\documentclass[aps, pre, twocolumn]{revtex4}
\usepackage{graphics}
\newcommand{\expec}[1]{\langle {#1} \rangle}

\begin{document}

\title{Dynamics of Nucleation in the Ising Model}

\author{Albert C. Pan and David Chandler}

\affiliation{Department of Chemistry, University of California, 
Berkeley, CA 94720-1460}

\date{\today}

\begin{abstract}
Reactive pathways to nucleation in a three-dimensional Ising model at 60\% of the
critical temperature are studied using transition path sampling of
single spin flip Monte Carlo dynamics.
 Analysis of the transition state ensemble (TSE) indicates
that the critical nuclei are rough and anisotropic.  The TSE,
projected onto the free energy surface 
characterized by cluster size,
$N$, and surface area, $S$, indicates the significance of other
variables in addition to these two
traditional reaction coordinates for nucleation.  The transmission coefficient, $\kappa$, along
$N$ is $\kappa \approx 0.35$, and this reduction of
the transmission coefficient from unity is explained in terms of the stochastic
nature of the dynamic model.  
   
\end{abstract}

\maketitle

\section{INTRODUCTION}
\label{intro}

This paper presents a new application of transition path
sampling\cite{FirstTPS, ShootingTPS, TPSAnnRev, TPSREV}, 
namely to nucleation of bulk phase transitions\cite{OxtobyReview}.  It should be of interest 
to those concerned with computational techniques devoted to rare 
transitions between metastable states, as well as to those 
interested in nucleation theory.  The application focuses on the 
simplest example of nucleation, that of a supercooled Ising model. 
We are not the first to carry out numerical simulations of nucleation 
in the Ising model.  For example, see \cite{Binder, LebowitzII, LebowitzIV,
 HeermannPRL,
  StaufferStatPhys,StaufferJCP, StaufferJMP, Landau, BeattyI, BeattyII}.   Our work is 
distinguished from these earlier studies in that we focus on the
statistics of an ensemble of reactive pathways to nucleation.
 We are also not the first to use 
transition path sampling to study nucleation of a bulk phase 
transition.  Zahn, for instance, has used this technique to study 
atomistic models undergoing solid-solid transitions\cite{Zahn2, Zahn}.  That work 
succeeded at harvesting typical nucleation pathways and examples of 
transition states for specific molecular systems.  In contrast, our 
focus in this paper is on generic issues raised by recent experiments 
and simulations\cite{Balsara1, Balsara2, Weitz, Frenkel3, Frenkel1,
  Frenkel2}, issues that suggest the importance of deviations 
from classical nucleation theory due to fluctuations\cite{Binder4, Binder}.

The thermodynamics of nucleation are
 thought to be governed
by a competition between two effects in
 the growing nucleus ---
an unfavorable contribution from the formation of a surface and a
 favorable contribution from nucleating the stable phase:
\begin{equation}
\label{CNT}
\Delta G(N) = - N |\Delta \mu | + N^{\frac{2}{3}}\gamma.  
\end{equation}
Here, $N$ is the number of particles in the growing nucleus (assumed to be 
spherical), $\Delta \mu$ is the chemical potential difference between
the two phases and $\gamma$ is the surface tension (assumed to be that
of
an infinite planar interface).  
Equation \ref{CNT} assumes that the free energy of this non-equilibrium
process depends on the size of the growing nucleus as the one
relevant variable (i.e. reaction coordinate) that controls its
progress: small nuclei tend to
shrink due to their large surface area-to-volume ratios while
sufficiently large nuclei tend to grow as the bulk free energy
dominates.  The transition state, or critical nucleus, then sits atop this free energy
barrier between the undercooled and stable phases.  This picture is
found in any theory that relates the nucleus size to a single reaction
coordinate.  In general, nucleation, as with all non-equilibrium
process, can involve many degrees of freedom\cite{Langer, Binder3} and may not be faithfully described by
one or even a small handful of coordinates\cite{Alanine, TPSNACL}.  

We investigate to what extent the simplest picture
holds.  Our analysis involves reversible work calculations,
committor distributions, transmission coefficients and, most
importantly, the statistics of an ensemble of reactive
trajectories.  We find that the traditional coordinate, $N$, provides
a reasonable approximation to the reaction coordinate.  But other
variables in addition to $N$ and also cluster surface area, $S$, are
required for a quantitative treatment.  We also find that the critical nuclei in the transition state
ensemble are rough and anisotropic as seen recently in
experiments on colloidal and polymer systems\cite{Weitz, Balsara1,
  Balsara2}.  

\section{MODEL and simulation details}
\label{model}

Our system is the nearest neighbor Ising model on a cubic lattice with
the Hamiltonian:
\begin{equation}
H = -J\sum_{\expec{ij}} s_i s_j + h\sum_i s_i
\end{equation}
where $J$ ($>0$) is the coupling constant, $h$ is the magnetic
field and $s_i$ is a spin variable that can either be $1$ or
$-1$.  The bracketed sum over $i$ and $j$
denotes a restriction to nearest neighbor pairs.  In
these simulations,  the temperature is 60\% of the critical temperature in
units of $J/k_B$ and the field $h = 0.55$ in units of $J$.  The lattice has 32 spins 
on an edge with
periodic boundary conditions and is propagated using single spin flip Metropolis Monte Carlo with random site
selection.  Time is measured in units of sweeps.

The temperature 0.6 $T_c$ is about 20\% above the roughening
temperature of the three dimensional Ising model, $T_R$\cite{Weeks}.
One anticipates that below $T_R$, the nuclei will tend to be cubic 
with relatively flat interfaces\cite{StaufferJMP,
  StaufferJCP}, while above $T_R$, nuclei will tend to be isotropic
and rounded\cite{RottmanWortis}.  This latter regime is appropriate for the study of
liquid-vapor equilibrium.  Using eqn. \ref{CNT}, we can anticipate the typical critical
cluster size assuming a spherically isotropic nucleus.  
Taking $|\Delta\mu| = 2h$ (due to the usual connections
between the Ising spin system and a lattice gas) and $\gamma \approx
2J$ (assuming the zero temperature value of $\gamma$ in the Ising model) 
implies
a critical cluster size, $N^* \sim 200$ with a free energy barrier of
about 40 $k_BT$.  In fact, the exact numerical analysis
gives $N^* = 115$ with a corresponding free energy barrier of 18
$k_BT$ (see below).  In any case, the size of the critical nucleus is
much smaller than our system size and the height of the free
energy barrier indicates that nucleation is a rare event.

\begin{figure} 
\resizebox{\columnwidth}{!}{
\includegraphics*{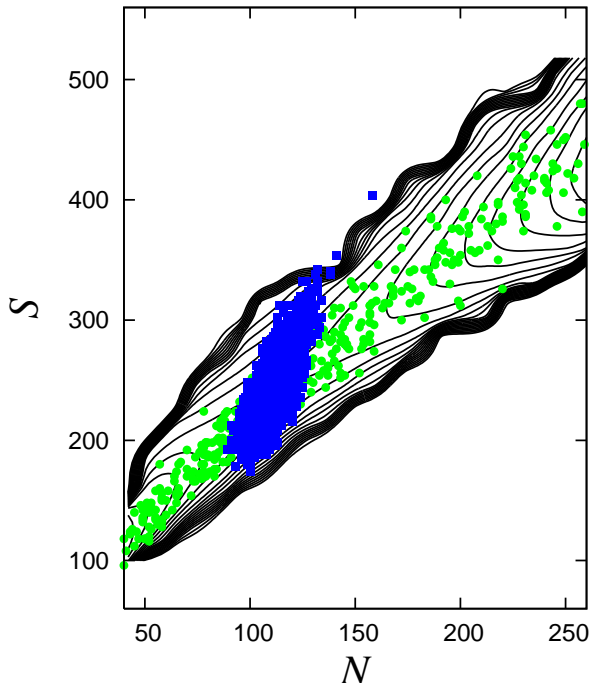}
}
\caption{Contour plot of $\Delta G(N, S)/k_BT$, the free energy of a
  nucleus as a function of its size and surface area. The contour
  lines are in gradations of 1 $k_BT$. The green circles show points
  visited by eight typical trajectories projected onto the $N$-$S$
  plane.  The blue squares show members of the transition state
  ensemble projected in the same way.}
\label{2Dfree}
\end{figure}
   
\section{Statistcal Analysis of an ensemble of reactive trajectories}
\subsection{Transition path sampling}
\label{tps}

In order
to study the dynamics of nucleation without being biased by a
particular choice of reaction coordinate, many trajectories of the
nucleation 
event need to be obtained without reference
to any specific coordinate.  For nucleation, such a calculation
might seem difficult to carry out because the process is a rare event.
The difficulty is overcome, however, with transition path sampling (TPS)\cite{TPSREV, TPSAnnRev}.     
In a straightforward simulation, a large majority of computational time is
spent simulating the undercooled state even though the nucleation
event of interest is fleeting
(see, for instance, \cite{NatureIce}).  In contrast,
TPS allows 
 exclusive sampling of the reactive portion of the trajectory.
It employs a Monte Carlo walk in the space of reactive trajectories to
harvest multiple examples of the rare event without wasting
computational time simulating the metastable state.  Moreover, no
reaction coordinate is required {\it a priori}.

\begin{figure}[b] 
\resizebox{\columnwidth}{!}{
\includegraphics*{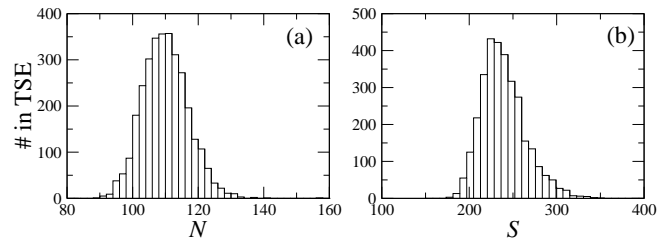}
}
\caption{Distributions of cluster sizes (a) and surface areas (b) in the transition state
  ensemble.  The average values from the two histograms are $N = 110$ and $S=241$. }
\label{ClusterSize}
 
\end{figure}

We apply TPS to nucleation in the Ising model and sample paths
connecting
nucleated and undercooled states defined by the characteristic
functions $h_A(q)$ and $h_B(q)$:
\begin{equation}
\label{ha}
h_A(q) = \left\{
	\begin{array}{cc}
	1, & N(q)<N_A\\
	0, & N(q)>N_A
	\end{array} 
	\right.
\end{equation}
\begin{equation}
\label{hb}
h_B(q) = \left\{
	\begin{array}{cc}
	0, & N(q)<N_B\\
	1, & N(q)>N_B
	\end{array} 
	\right.
\end{equation}
Here, $q = (s_1, s_2, ... , s_i, ...)$ denotes a particular
configuration of
the lattice and $N(q)$ returns the
size of the largest cluster in that configuration.  The limits, $N_A$
and $N_B$ are far removed from the
transition state region.  In other words, $h_B(q)$
gives a signal if a configuration is in the product region and $h_A(q)$
gives a signal if a configuration is in the reactant region.  In this
calculation,
$N_A = 26$ and $N_B = 260$ are chosen such that the free energy barrier
in FIG. \ref{umbrella} (see below) separating the two basins
exceeds 10 $k_BT$.  This ensures that once a configuration finds itself in either
the reactant or product region, it remains there for times much longer
than the molecular relaxation time and that the reactant and product basins
do not overlap.  Since $N$ may not necessarily be an adequate
reaction coordinate, we verify this latter condition by separate
simulations of configurations starting in both regions. 

\begin{figure} 
\resizebox{\columnwidth}{!}{
\includegraphics*{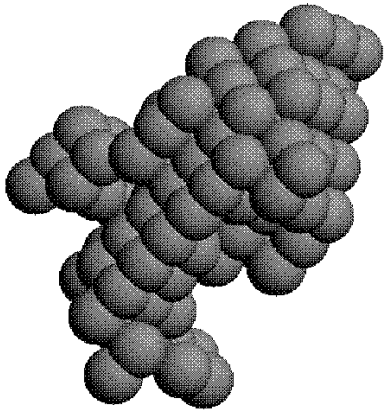}
}
\caption{A characteristic example from the transition state ensemble.
The spheres represent nucleated spins on the cubic Ising lattice.}
\label{CriticalNucleus}
 
\end{figure}

In these simulations, trajectories 150 time units in length are
sampled with the shooting algorithm\cite{ShootingTPS, TPSREV}: a time slice is chosen at random from a
 trial trajectory and then new forward and backward paths are
 generated using
 the underlying dynamics of the system.  
The newly generated trajectory is accepted if it connects the reactant
 and product regions defined by the characteristic functions $h_A(q)$
 and $h_B(q)$.
 Since the dynamics of this system are stochastic, a new path can be
 generated
by simply shooting one direction at a time, as the forward and backward
transition probabilities are equal\cite{TPSREV}.  This increases the
 acceptance probability.  A complete shooting move is defined as two such moves.
 We relax an initial trajectory\cite{inittraj} with 25,000 moves and then 1,000 independent
trajectories
are harvested, one every 100 moves.  Convergence and adequate choice
of path length are verified by calculation of $\langle h_B[q(t)]
\rangle_{AB}$, the characteristic function $h_B(q)$ along a reactive
trajectory averaged over the transition path ensemble (not shown).
The fact that this quantity reaches the linear regime implies that our
path length is long enough to sample typical barrier crossing behavior\cite{TPSREV}.  

One often imagines that the most important reaction coordinate
describing nucleation is the size of the growing cluster, $N$, as
alluded to in the Introduction.  Another relevant, but secondary, reaction coordinate
which is also considered is the cluster's surface area, $S$.  In FIG. \ref{2Dfree}, 
we project paths obtained from TPS onto a contour plot of
the $N$-$S$ free
energy (green circles). 

\begin{figure}[b] 
\resizebox{\columnwidth}{!}{
\includegraphics*{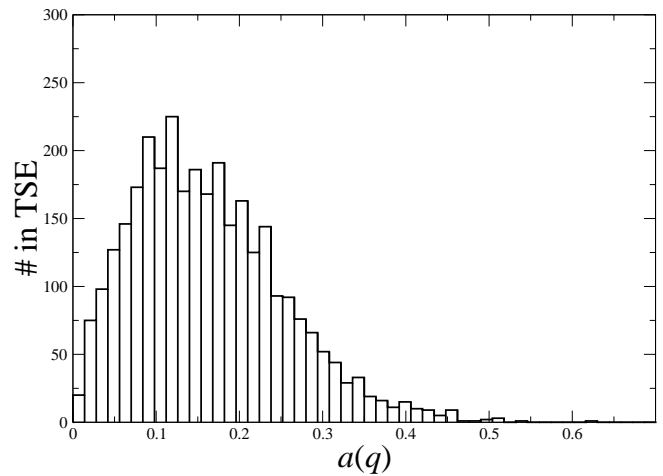}
}
\caption{The distribution of $a(q)$, a function of the
  principal moments
  of inertia of the nuclei (see eqn. \ref{peq}), in the transition state
  ensemble.}
\label{PrincipalMoment}
 
\end{figure}

The free energy surface in FIG. \ref{2Dfree} is determined using
umbrella sampling with hard wall constraints\cite{ChandlerBook}.  For this and all subsequent calculations, a set of spins
is considered a nucleus of size $N$ if each spin in the set is a nearest neighbor of at
least one other spin in the set.  The surface area, $S$,
 of such a cluster is the number of exposed faces.  The umbrella sampling
 windows constrain the size and surface area of the {\em largest} cluster in a given
configuration of the lattice.  For $N > 20$, it is highly unlikely that
there is more than one cluster of that size in any given
configuration of the full system.  Approximately 200
overlapping windows of average size $\Delta N = 12$ by $\Delta S =
40$ are used.  In each window, an initial configuration is
equilibrated for 50,000 sweeps and statistics are taken over a
subsequent run of 200,000 sweeps.  The windows are linked together
with multiple histograms generalized to two dimensions\cite{FrenkelandSmit}.

\subsection{The transition state ensemble}
\label{tse}

A configuration is considered a member of the transition state
ensemble (TSE) if half of the new trajectories initiated from it cause the
nucleus to shrink and the other half cause it to grow.  From the 1,000
trajectories acquired through transition path sampling, we found approximately
3,200 members of the TSE. 

To determine transition states with a minimum of computational
effort, we check each configuration along a
reactive trajectory 
in the following way\cite{TPSREV}.  (1) If a
configuration is in either the reactant or product region, it is
rejected as a transition state candidate straight away.  (2) If a
configuration is not in the reactant or product region, a series of
additional trajectories are initiated from that configuration and
$p_B$, the ratio of paths which end up in region $B$ to the total
number of paths initiated, is calculated after 11, 14, 17, 20, 25, 30, 35, 42 and 49
trajectories\cite{Tom}. 
After the 49th trajectory, additional trajectories are generated up
to a maximum of 100 trajectories, and $p_B$ is calculated after every trajectory.
The configuration is rejected as a transition state candidate if
$p_B$ falls outside the $95\%$ confidence interval around $p_B = 0.5$
at any point.  Finally, (3) if $p_B = 0.5$ within $95\%$ confidence
after
100 trajectories, then that configuration is accepted as a member of
the transition state ensemble.  

FIG. (\ref{2Dfree}) juxtaposes projections onto the $N$-$S$ 
plane of representative trajectories and the TSE 
with the corresponding free energy surface. The 
comparisons indicate that $N$ and $S$ capture much 
of the mechanism for nucleation. The TSE is not 
perpendicular to the $N$ axis, showing that $S$ as 
well as $N$ is important to the the mechanism of 
nucleation.  The comparisons also show that other 
variables in addition to $N$ and $S$ play significant 
roles in the mechanism.  In particular, the 
orientation of the projected TSE is far from that 
expected from the saddle in the free energy 
surface.  Further, the projected TSE has a 
significant width.

Further analysis of the TSE shows that the critical nuclei are rough
and anisotropic.  The distribution of cluster sizes in
the transition state ensemble is shown in FIG. \ref{ClusterSize}(a).  
A typical critical nucleus taken from the transition state ensemble is
shown in FIG. \ref{CriticalNucleus}.  The anisotropy and roughness
of this nucleus are characteristic of the transition state ensemble. 
Quantitative measurements of this fact are shown in FIG.
\ref{ClusterSize}(b) and FIG. \ref{PrincipalMoment}.  
FIG. \ref{ClusterSize}(b) shows the distribution of surface areas in
the transition state ensemble.  
The average surface area is 241 compared
to an average cluster size of 110 (FIG. \ref{ClusterSize}).  This observed
average surface area is almost 30\% larger than would be expected of a
compact spherical cluster of 110 particles indicating that the critical nuclei are, on
average, extremely rough\cite{sarea}.  

FIG. \ref{PrincipalMoment} shows the distribution of the anisotropy
function,
\begin{equation}
\label{peq}
a(q) = \frac{I_1(q)}{I_2(q)} - 1,
\end{equation}  
where $I_1(q)$ and $I_2(q)$ are the largest and second largest,
respectively, principal moments of inertia.  For a completely
isotropic structure, $a(q) = 0$.  
The deviation from zero indicates anisotropy.  In contrast, the
equilibrium average crystal shape
for a nearest neighbor three-dimensional Ising model above the
roughening transition is isotropic and rounded\cite{RottmanWortis}.    

In the following sections, we contrast these results obtained from the
statistical analysis of an ensemble of reactive trajectories with
those obtained from more conventional methods.

\begin{figure} 
\resizebox{\columnwidth}{!}{
\includegraphics*{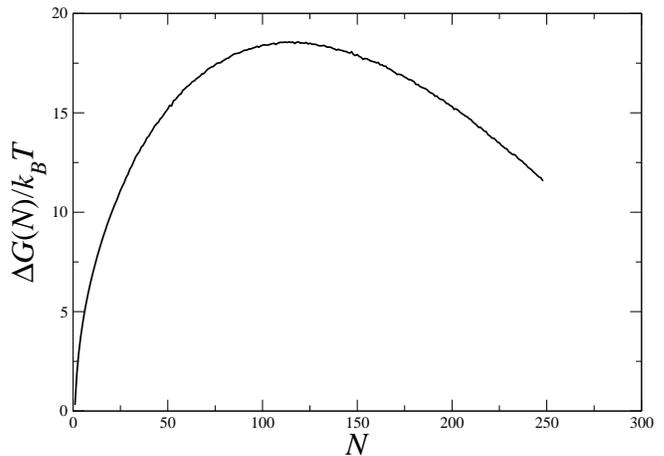}
}
\caption{The free energy of a growing cluster in the three-dimensional Ising model
at $T=$ 0.6 $T_c$ and $h =$ 0.55 $J$.}
\label{umbrella}
\end{figure}

\section{Cluster size as reaction coordinate}
\subsection{Reversible work of cluster formation}

FIG. \ref{umbrella} shows the free energy $\Delta G(N)$ where, within
an additive constant, $\Delta
G(N) = -k_BT\sum_S\exp[-\Delta G(S, N)]$.  Qualitatively, we see that the computed curve resembles the curve
predicted by equation \ref{CNT}.  The maximum occurs at $N^*=115$
monomers where $\Delta G(N^*)$ is within a small fraction of $k_BT$
from the free energy at $N=110$ monomers,
the average
of $N$ in the TSE.

The free energy in FIG. \ref{umbrella} is calculated using umbrella
sampling with hard wall constraints, and different umbrella windows are linked together using the
multiple histogram method.  Approximately 30
overlapping windows of size $\Delta N = 12$ are used.  In each window, an initial configuration
is equilibrated with 50,000 sweeps and then statistics are taken over
a run of 100,000 sweeps.  

\begin{figure}[b] 
\resizebox{\columnwidth}{!}{
\includegraphics*{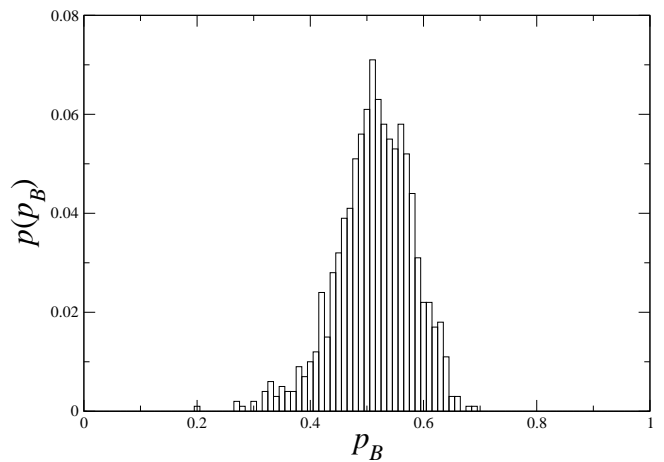}
}
\caption{Committor distribution with cluster size $N$ constrained to $N^*$.  
The distribution is peaked around $p_B=50\%$ indicating that $N$ is a reasonable
reaction coordinate for this process.}
\label{committor}
\end{figure}

\subsection{Committor distribution}
\label{committors}

The extent to which $\Delta G(N)$ provides an adequate indication
of the dynamics can be determined by calculating the probability that
configurations constrained to have a nucleus of size $N^*$ will either 
grow or shrink.  If $N$ is indeed a good reaction coordinate, a
configuration
with an $N^*$ sized nucleus should be just as likely to grow as to
shrink.  
In this case, the probability distribution, also called a committor
distribution, would be peaked around 50\%\cite{TPSREV, TPSAnnRev}.  

The distribution in FIG. \ref{committor} represents the results of
such a calculation where $p_B$ once again denotes the probability of a
configuration 
ending up in the product region.
A set of 1,000 
independent configurations is drawn from the
ensemble of 
configurations constrained to have $N=N^*$ and 200 separate
trajectories are then run for each configuration.  The final
configuration
of these trajectories is then judged by the characteristic
function $h_B(q)$ (eqn. \ref{hb}).

The fact that the committor distribution for $N$ constrained to $N^*$
is peaked around 50\% indicates that it is a reasonable
approximation to the reaction
coordinate for nucleation.  If this were not the case, 
we would expect a different distribution (see, for
example, \cite{TPSNACL}).  The spread in
the distribution, however, indicates that coordinates
other than $N$ are still involved albeit in a secondary way.  

\subsection{The transmission coefficient along N}
\label{rflux}

We calculate the transmission coefficient, $\kappa$, via the reactive
flux method\cite{rflux, ChandlerBook, Frenkel1}.  The value of
$\kappa$ is then given by the plateau of the normalized reactive flux correlation
function:
\begin{equation}
\label{kappaeq}
k(t) = \frac{\expec{\dot{N}(0)\theta[N(t)-N^*]}'}
{\expec{\dot{N}(0) \theta[\dot{N}(0)]}'}
\end{equation}
where $\theta$ is the Heaviside step function, $N(t)$ is shorthand for
$N[q(t)]$, and the primed angled brackets indicate an
equilibrium average with $N(0)$ constrained to $N^*$.  In other words, we constrain our
ensemble of initial states to be at the maximum, $N^*$, of $\Delta G(N)$
(FIG. \ref{umbrella}). 

\begin{figure} 
\resizebox{\columnwidth}{!}{
\includegraphics*{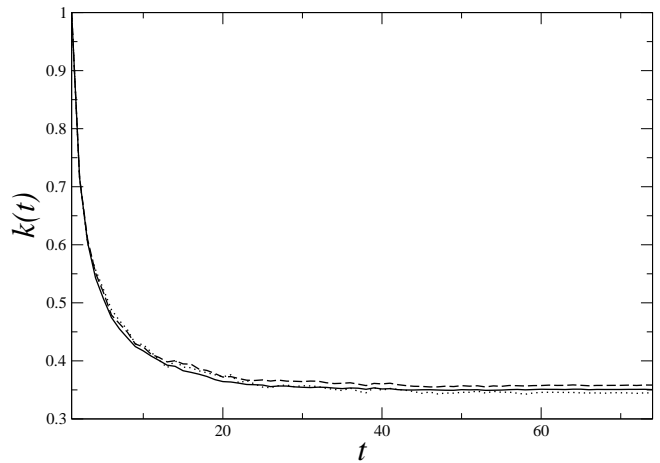}
}
\caption{Plot of $k(t)$ (eqn. \ref{kappaeq}), where time is measured
  in Monte Carlo sweeps.  The dotted, dashed and solid lines are averages over 10,000, 30,000
  and 80,000 trajectories, respectively.}
\label{kappa}
\end{figure}

A plot of $k(t)$ is given in
FIG. \ref{kappa}.  A plateau value of $\kappa < 1$ along N is an indicator of recrossings due to friction in
the barrier region.  Here, $\kappa \approx
0.35$.  We argue below that the friction in our system is mainly a
manifestation of the stochastic dynamics in the Monte Carlo trajectory. 

For the calculations in FIG. \ref{kappa}, a set of independent configurations is drawn from the
ensemble of 
configurations constrained to have $N=N^*$ and trajectories are run
beginning from these configurations.  The reactive flux correlation
function, $k(t)$, is then calculated as an average over these
trajectories.  The initial velocity of the reaction coordinate,
$\dot{N}(0)$, is taken to be the finite difference, $N(1) - N(0)$.

\subsection{Friction from stochastic dynamics}

The friction which reduces the value of the transmission coefficient
in this case can be attributed mostly to the diffusive
nature of a random walk on a relatively flat barrier top.  To
illustrate this idea, we consider a random walk beginning at the top
of a free energy barrier
and calculate its transmission coefficient.  Here, we assume that the random walker makes
uncorrelated steps of typical length $\delta N$ and is committed to a basin
once it has traveled a distance $l$ corresponding to when the free
energy has changed by $\sim$ 1 $k_BT$ relative to the barrier top.
The length $l$ therefore depends on the curvature of the barrier near its
maximum.

\begin{figure} 
\resizebox{\columnwidth}{!}{
\includegraphics*{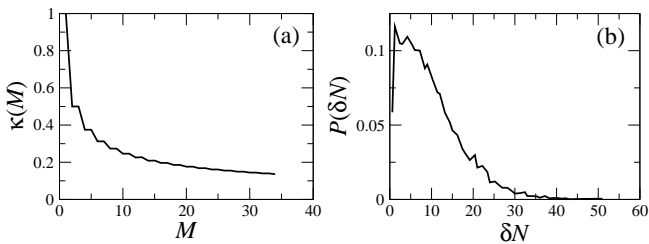}
}
\caption{(a) The transmission coefficient $\kappa$ for a random walker
as a function of the number of steps $M$ required for it to fall off
the barrier top.  (b) Probability of observing a change in the nucleus
size, $\delta N$ after one sweep of the lattice in the region 1 $k_BT$
around the barrier top in FIG. \ref{umbrella}, $N \in [80, 160]$ (averaged
over 500 trajectories).}
\label{pdn}
\end{figure}

The reactive flux correlation function in eqn. (\ref{kappaeq}) can be
thought of as a ratio of the average flux across the dividing surface
 of trajectories which end up in the product region to
the average flux across the dividing surface of trajectories with an
initial
positive flux (i.e. toward the product region). 
For a random
walk process, these quantities can be evaluated analytically from
a straightforward application of binomial statistics.  The quantity
of interest is the value of the transmission coefficient $\kappa(M)$
where $M \sim l/\delta N$ is the number of steps required to fall a
distance 1 $k_BT$ in free energy from the barrier top.  If $M = 1$,
there are only two possible trajectories --- one with positive initial
flux which is trapped on the positive side of the dividing surface and one with initial negative
flux which is trapped on the negative side.  In this case, we see that
$\kappa(1) = 1$.  Similarly for $M=2$, out of four possible trajectories,
one trajectory is trapped on the positive side of the dividing surface with
initial positive flux, one trajectory is trapped on the negative side with
initial negative flux, and the other two trajectories end up back at
the dividing surface, one with positive initial flux and one with
negative initial flux.  In this case, $\kappa(2) = 1/2$.  In general,
we see that the denominator of $\kappa(M)$ is half
of all possible trajectories of length $M$ and that the numerator is,
within
the subset of trajectories of length $M$ which end up in the product region, the
number with a positive initial flux minus the number with a negative
initial flux.        
Therefore, for arbitrary (even) $M$, we have:
\begin{equation}
\label{binkappa}
\kappa(M) = n(M)/d(M)
\end{equation}
where,
\begin{equation}
d(M) = \frac{1}{2}\sum_{r = 0}^{M}\limits {M \choose r},
\end{equation}
and, for even $M$,
\begin{equation}
n(M) = 1 + \sum_{r = 1}^{M/2-1}\limits\left[ {{M-1}\choose r} - {{M-1}\choose {r-1}}\right]
\end{equation}
with ${i \choose j} = i!/(i-j)!j!$, as usual.  For odd $M$, the result is the
same except the upper limit in the sum giving $n(M)$ is changed to
$(M-1)/2$.
$\kappa(M)$ is plotted
in FIG. \ref{pdn}(a).  For the
transmission coefficient of nucleation, the distance to 1 $k_BT$ from
the barrier top is $l \sim 45$ and the typical random walk step size
is $\delta N \sim 11$.  The latter result can be arrived at by
considering the typical size of fluctuations of a nucleus of size
$N^*$.  In this case, $\delta N \sim \sqrt{N^*} \sim 11$.
Alternatively, one can compute the probability of observing a change,
$\delta N$, in the nucleus size after one sweep.  This probability, depicted in
FIG. \ref{pdn}(b), shows that $\delta N \sim 11$ is a reasonable
estimate.  These numbers imply that the typical random walk step
 size is approximately 4 which gives $M \sim (l/\delta
 N)^2 \sim 16$.  This leads
to an estimate of $\kappa \approx 0.20$.  Considering the rough nature
of the approximation, this result is close to
the simulation result of $\kappa \approx 0.35$ indicating that the
stochastic nature of the dynamics is really playing the major role in 
determining the value of the transmission coefficient along $N$.  

Zeldovich was the first to write down an analytic expression for the
transmission coefficient, $Z$, for nucleation\cite{FrenkelZeldo}: 
\begin{equation}
Z = \left\{\frac{1}{2\pi}\left[\frac{\partial^2(\Delta G(N)/k_BT)}{\partial N^2}\right]_{N^*}\right\}^{1/2}
\end{equation}
This factor, a measure of the barrier width, is an indication
of the diffusive nature of nucleation dynamics.  In a more modern context, we see that $Z$
is proportional to the high friction limit of Kramers expression for
$\kappa$\cite{HanggiTalknerBorkovec}.  The Zeldovich factor
has units of $1/N$ and therefore depends also on the size of the
nucleus' typical fluctions.  In our system, $Z \approx 0.013$, which,
when multipled by $\delta N \sim 11$, gives a reasonable estimate of $\kappa
\approx 0.14$.  

An atomistic simulation study of liquid-gas nucleation gives a value
of $\kappa$ two orders of magnitude smaller than 0.35\cite{Frenkel2}.  The coarse
grained dynamics used in the current study takes much larger steps in
configuration space
than atomistic dynamics.  The corresponding $M$ for the atomistic
dynamics
is thus much larger than that which we associate with the Monte Carlo
random walk.  The much larger value of $M$ can explain the much
smaller
value of $\kappa$, as eqn. \ref{binkappa} shows.

The authors thank Pieter Rein ten Wolde for discussions and advice on
this work, which has been supprted by the US National Science
Foundation.  A.C.P. is a NSF Graduate Research Fellow. 

\providecommand{\refin}[1]{\\ \textbf{Referenced in:} #1}


\begin{thebibliography}{10}
\bibitem{inittraj}
The initial trajectory was generated by taking a configuration with $N=N^*$
  generated during umbrella sampling and running trajectories until a path was
  found connecting the reactant and product basins. Even though this path may
  have been highly unphysical, subsequent relaxation through shooting moves,
  analogous to Monte Carlo relaxation of an intial state before sampling,
  equilibrated the path.
\bibitem{sarea}
The minimal surface area for a nucleus of size $N$ was calculated by creating a
  compact spherical cluster with $N$ monomers on a three dimensional Ising
  lattice and determining its surface area.
\end{thebibliography}
\end{document}